# WHU-Hi: UAV-borne hyperspectral with high spatial resolution (H$^2$) benchmark datasets for hyperspectral image classification.

Xin Hu, Yanfei Zhong*, Chang Luo, Xinyu Wang

*Wuhan University, Wuhan, China*

{whu_huxin, zhongyanfei, luochang, wangxinyu,} @whu.edu.cn

## Abstract

*Abstract*—Classification is an important aspect of hyperspectral images processing and application. At present, the researchers mostly use the classic airborne hyperspectral imagery as the benchmark dataset. However, existing datasets suffer from three bottlenecks: (1) low spatial resolution; (2) low labeled pixels proportion; (3) low degree of subclasses distinction. In this paper, a new benchmark dataset named the Wuhan UAV-borne hyperspectral image (WHU-Hi) dataset was built for hyperspectral image classification. The WHU-Hi dataset with a high spectral resolution (nm level) and a very high spatial resolution (cm level), which we refer to here as H2 imager. Besides, the WHU-Hi dataset has a higher pixel labeling ratio and finer subclasses. Some start-of-art hyperspectral image classification methods benchmarked the WHU-Hi dataset, and the experimental results show that WHU-Hi is a challenging dataset. We hope WHU-Hi dataset can become a strong benchmark to accelerate future research.

## 1. Introduction

Hyperspectral imaging combines the traditional two-dimensional imaging technology with spectral technology, and records tens or even hundreds of continuous narrow-band information of each pixel when detecting the spatial information of the measured object. At present, the rich hyperspectral data source has been provided by satellites, airplanes, and now unmanned aerial vehicle (UAV) observation platforms[1, 2]. When compared with space-borne and manned aircraft borne hyperspectral imaging systems, a new Earth observation platform—unmanned aerial vehicle (UAV)-borne hyperspectral systems—can simultaneously acquire hyperspectral imagery with high spatial and spectral resolution (H$^2$ imagery)[3].

Classification is an important aspect of HSIs the processing and application of hyperspectral images (HSIs),, and its ultimate goal is to assign a specific class to each pixel in the image. The accurate classification of HSIs had been applied extensively in various applications, such as land use monitoring, precision agriculture, and urban planning. At present, as shown in 1 to 8 of Table 1, the researchers mostly use the classic airborne hyperspectral imagery as the benchmark dataset. However, existing benchmark datasets suffer from three bottlenecks. (1) Low spatial resolution. The resolution is mainly from the meter level to the ten-meter level. (2) Low labelled pixels proportion. The labelling ratio is generally less than 50%, and the number of labelled pixels is also small. (3) Low category fineness. For some datasets, such as Pavia University, Pavia Center, the spectra between different classes are quite different and easy to distinguish. These bottlenecks limit the study of the finer classification of ground features, and a new hyperspectral classification dataset with higher spatial resolution, higher proportion of labelled pixels and more categories is needed. To solve these problems, we built the WHU-Hi UAV-borne hyperspectral dataset, which is the first open-source UAV-borne H² dataset. The WHU-Hi dataset is made up of the WHU-Hi-LongKou, WHU-Hi-HanChuan and WHU-Hi-HongHu datasets, as shown in 9 to 11 of Table 1, which has the following characteristics. (1) Very high spatial resolution. The resolution of WHU-Hi dataset is from the decimeter level to the centimeter level. (2) Very high labeled pixels proportion. For each dataset, the proportion of the labeled pixel is more than 69% and labeled pixels' number exceeds 200,000. (3) Abundant ground feature types. All the datasets were acquired in farming areas with various crop types in Hubei province, China, and WHU-Hi-HongHu dataset contains up to 18 types of crops.

Table 1: open-source datasets in HSIs classification

| No. | Dataset | Number of bands | Spatial resolution | Number of classes | Image size | Label ratio | Sensor | Time |
|---|---|---|---|---|---|---|---|---|
| 1 | Botswana | 145 | 30 | 14 | 1476×256 | 0.86% | Hyperion | 2001-2004 |
| 2 | Kennedy Space Center | 176 | 18m | 13 | 512×614 | 1.66% | AVIRIS | 1996 |
| 3 | Indian Pines | 224 | 16m | 16 | 145×145 | 48.75% | AVIRIS | 1992 |
| 4 | Salinas | 224 | 3.7m | 16 | 512×217 | 48.72% | AVIRIS | 1998 |
| 5 | Houston | 144 | 2.5m | 15 | 349×1905 | 2.26% | CASI 1500 | 2012 |
| 6 | Pavia University | 103 | 1.3m | 9 | 610×340 | 17.43% | ROSIS | 2003 |
| 7 | Pavia Center | 102 | 1.3m | 9 | 1096×715 | 18.91% | ROSIS | 2003 |
| 8 | Tea Tree dataset | 80 | 2.25m | 10 | 348×512 | 30.16% | PHI | 1999 |
| 9 | WHU-Hi-LongKou | 270 | 0.463 m | 9 | 550×400 | 92.97% | Nano-Hyperspec | 2018 |
| 10 | WHU-Hi-HanChuan | 274 | 0.109 m | 16 | 1217×303 | 69.84% | Nano-Hyperspec | 2016 |
| 11 | WHU-Hi-HongHu | 270 | 0.043 m | 22 | 940×475 | 86.61% | Nano-Hyperspec | 2017 |

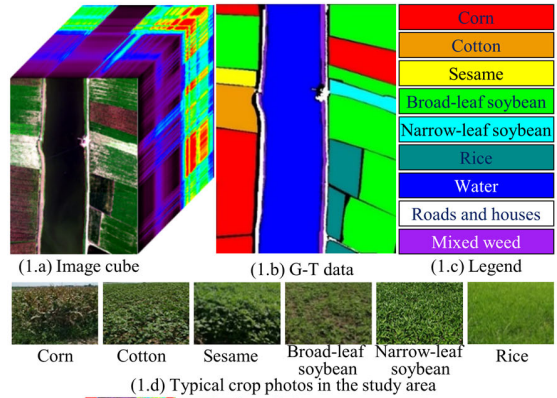
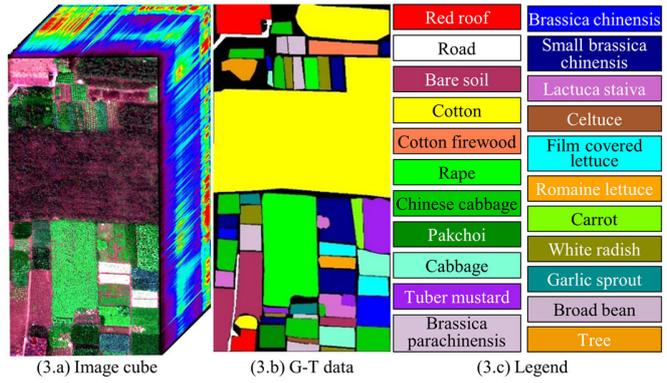
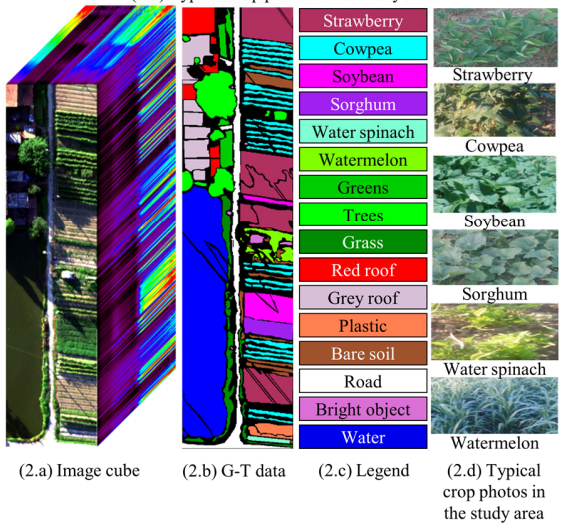
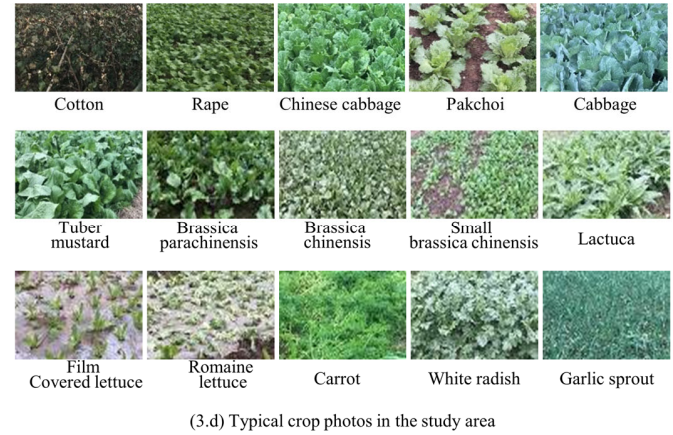
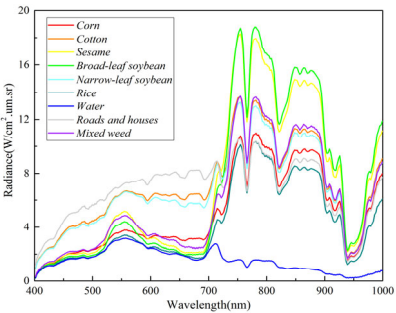
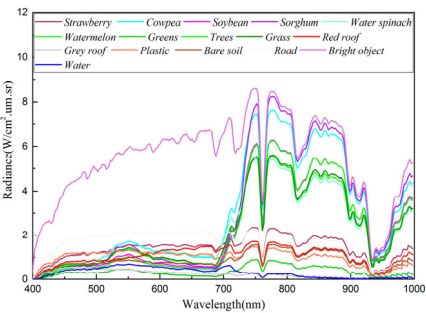
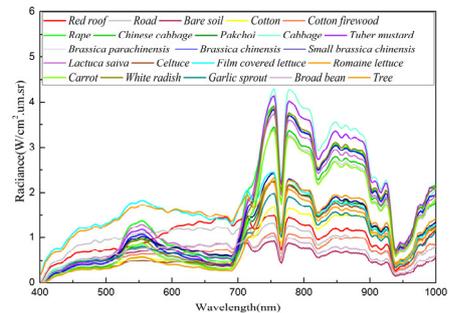

Fig. 1. Wuhan UAV-borne H² imagery (WHU-Hi) dataset. (1) The WHU-Hi-LongKou dataset. (2) The WHU-Hi-HanChuan dataset. (3) The WHU-Hi-HongHu dataset. (G-T: ground truth.).

## 2. WHU-Hi Dataset

As shown in Fig. 1, The WHU-Hi dataset was acquired using a Headwall Nano-Hyperspec sensor that was equipped on a UAV platform, in farming areas with various crop types in Hubei province, China. The UAV-borne hyperspectral data preprocessing included radiometric calibration and geometric correction, which were completed in the Hyperspec software provided by the instrument manufacturer. For the radiometric calibration, the raw digital number (DN) value was converted into radiance value by the calibration parameters of the sensor. The geometric correction was then undertaken, based on the collinear equation and the position and attitude information recorded by the global positioning system (GPS)/inertial measurement unit (IMU) module.

The scene of WHU-Hi-LongKou is a simple agricultural region, which was acquired from 13:49 to 14:37 on July 17, 2018, in Longkou Town, Hubei province, China, with an 8-mm focal length Headwall Nano-Hyperspec imaging sensor equipped on a DJI Matrice 600 Pro (DJI M600 Pro) UAV platform. The study area contains six crop species: corn, cotton, sesame, broad-leaf soybean, narrow-leaf soybean, and rice. The UAV flew at an altitude of 500 m, the size of the imagery is 550 × 400 pixels, there are 270 bands from 400 to 1000 nm, and the spatial resolution of the UAV-borne hyperspectral imagery is about 0.463 m.

The scene of the WHU-Hi-HanChuan dataset is a rural-urban fringe zone, which was acquired from 17:57 to 18:46 on June 17, 2016, in Hanchuan, Hubei province, China, with a 17-mm focal length Headwall Nano-Hyperspec imaging sensor equipped on a Leica Aibot X6 UAV V1 platform. The study area contains seven crop species: strawberry, cowpea, soybean, sorghum, water spinach, watermelon, and greens. The UAV flew at an altitude of 250 m, the size of the imagery is 1217 × 303 pixels, there are 274 bands from 400 to 1000 nm, and the spatial resolution of the UAV-borne hyperspectral imagery is about 0.109 m. Notably, since the WHU-Hi-HanChuan dataset was acquired during the afternoon when the solar elevation angle was low, there are many shadow-covered areas in the image.

The scene of the WHU-Hi-HongHu dataset is a complex agricultural zone, which was acquired from 16:23 to 17:37 on November 20, 2017, in Honghu City, Hubei province, China, with a 17-mm focal length Headwall Nano-Hyperspec imaging sensor equipped on a DJI Matrice 600 Pro UAV platform. The study area is typical of the regions affected by land fragmentation, and is planted with 17 crop types, including cotton, rape, and cabbage. Notably, the region is planted with different cultivars of the same crop type; for example, Chinese cabbage/cabbage and brassica chinensis/small brassica chinensis. The UAV flew at an altitude of 100 m, the size of the imagery is 940 × 475 pixels, there are 270 bands from 400 to 1000 nm, and the spatial resolution of the UAV-borne hyperspectral imagery is about 0.043 m.

Table 2: The quantitative evaluation of the baseline methods for WHU-Hi dataset

| Dataset | | Accuracy | SVM | FNEA-OO | SVRFMC | SSAN | SSRN | pResNet | CNNCRF | SSFCN | FPGA |
|---|---|---|---|---|---|---|---|---|---|---|---|
| WHU-Hi-Longkou | 25 | OA | 91.54 | 98.37 | 98.18 | 86.90 | 91.74 | 97.78 | 97.31 | 87.70 | 95.67 |
| | | AA | 87.84 | 97.70 | 96.19 | 86.24 | 94.95 | 96.87 | 96.33 | 88.67 | 95.92 |
| | | Kappa | 0.8903 | 0.9806 | 0.9761 | 0.8581 | 0.8943 | 0.9710 | 0.9648 | 0.8438 | 0.9436 |
| | 50 | OA | 93.23 | 98.40 | 98.20 | 93.94 | 98.09 | 97.71 | 97.62 | 94.28 | 97.28 |
| | | AA | 92.38 | 98.05 | 97.05 | 92.65 | 96.85 | 98.04 | 96.55 | 95.14 | 97.46 |
| | | Kappa | 0.9124 | 0.9812 | 0.9780 | 0.9212 | 0.9750 | 0.9701 | 0.9688 | 0.9259 | 0.9675 |
| | 100 | OA | 94.96 | 98.59 | 98.37 | 94.44 | 99.02 | 98.70 | 98.91 | 94.60 | 99.17 |
| | | AA | 95.18 | 97.48 | 97.41 | 95.38 | 99.39 | 98.88 | 98.21 | 95.27 | 99.30 |
| | | Kappa | 0.9345 | 0.9815 | 0.9786 | 0.9279 | 0.9871 | 0.9830 | 0.9857 | 0.9300 | 0.9912 |
| WHU-Hi-HanChuan | 25 | OA | 61.80 | 67.75 | 69.05 | 74.80 | 76.19 | 82.28 | 86.94 | 73.44 | 89.28 |
| | | AA | 60.76 | 69.66 | 70.89 | 72.86 | 72.86 | 82.64 | 83.08 | 68.73 | 87.49 |
| | | Kappa | 0.5688 | 0.6345 | 0.6491 | 0.6955 | 0.7627 | 0.7963 | 0.8480 | 0.6949 | 0.8753 |
| | 50 | OA | 73.06 | 81.02 | 82.02 | 83.63 | 83.28 | 90.19 | 90.71 | 87.52 | 94.73 |
| | | AA | 67.89 | 78.98 | 79.63 | 82.73 | 80.48 | 89.30 | 89.78 | 84.01 | 94.66 |
| | | Kappa | 0.6901 | 0.7802 | 0.7919 | 0.8109 | 0.8028 | 0.8858 | 0.8917 | 0.8547 | 0.9385 |
| | 100 | OA | 77.61 | 85.63 | 86.53 | 88.63 | 89.82 | 93.32 | 93.95 | 89.75 | 97.83 |
| | | AA | 73.46 | 83.21 | 84.37 | 88.25 | 90.09 | 92.91 | 92.69 | 86.42 | 97.88 |
| | | Kappa | 0.7414 | 83.30 | 84.35 | 86.73 | 88.15 | 92.21 | 92.90 | 87.42 | 0.9747 |
| WHU-Hi-HongHu | 25 | OA | 66.66 | 85.09 | 84.84 | 72.82 | 81.17 | 86.89 | 84.84 | 82.92 | 91.33 |
| | | AA | 62.70 | 76.58 | 80.87 | 68.77 | 82.09 | 84.25 | 85.60 | 77.18 | 90.54 |
| | | Kappa | 0.6073 | 0.8127 | 0.8119 | 0.6810 | 0.7700 | 0.8365 | 0.8134 | 0.7864 | 0.8912 |
| | 50 | OA | 67.47 | 86.59 | 86.14 | 78.89 | 85.99 | 91.46 | 91.06 | 90.26 | 95.76 |
| | | AA | 66.12 | 80.75 | 84.87 | 74.75 | 88.36 | 91.44 | 91.12 | 86.81 | 96.49 |
| | | Kappa | 0.6116 | 0.8319 | 0.8287 | 0.7389 | 0.8281 | 0.8932 | 0.8881 | 0.8775 | 0.9467 |
| | 100 | OA | 73.55 | 88.83 | 89.86 | 87.34 | 91.29 | 95.32 | 93.74 | 94.26 | 97.45 |
| | | AA | 71.23 | 83.13 | 87.16 | 84.93 | 91.68 | 95.83 | 94.78 | 92.03 | 97.79 |
| | | Kappa | 0.6805 | 0.8590 | 0.8728 | 0.8415 | 0.8910 | 0.9412 | 0.9217 | 0.9219 | 0.9678 |

## 3. Benchmark

In order to establish a fair benchmark, we evaluated some state-of-the-art HSI classification methods under a unified experimental setting and data division conditions.

**Methods** After decades of development, HSI classification methods have evolved from pixel spectral classification methods based on statistical learning to deep learning classification methods considering global spatial spectrum information. In this paper, the different types of HSIs classification methods are used for experiments, including pixel spectral classification method (support Vector Machines (SVM)), the object-oriented classification method (the fractal net evolution approach (FNEA-OO)), the conditional random field-based classification method (the support vector conditional random fields classifier with a Mahalanobis distance boundary constraint (SVRFMC)[4]), the deep learning HSIs classification method based on spatial patching (the spectral-spatial attention network (SSAN[5]), the spectral-spatial residual network (SSRN)[6], deep pyramidal residual networks (pResNet)[7]), the combining CNN and CRF classification framework(CNNCRF)[3], the full convolution HSIs classification method (Spectral-Spatial Fully Convolutional Networks(SSFCN)[8], Fast Patch-Free Global Learning Framework (FPGA))[9].

**Settings** For each class, 25,50 and 100 labeled pixels were randomly selected for the model training, and the remaining pixels were used for the testing. To be more specific, the total number of training pixels was only 0.11%, 0.15%, and 0.10% of all the labeled pixels for the WHU-Hi-LongKou, WHU-Hi-HanChuan, and WHU-Hi-HongHu datasets, respectively. To quantitatively evaluate the experimental results, three evaluation indicators are used: the overall accuracy (OA), the average accuracy (AA), and the kappa coefficient (kappa).

**Analysis** The experimental results are shown in Table 2. For the traditional classification method, compared with the spectral classifier SVM, the classification accuracy of the spectral-spatial classifier FNEA-OO and SVRFMC has been greatly improved. And in the simple scene of WHU-Hi-LongKou, the FNEA-OO and SVRFMC methods have achieved better results under a small number of samples. Compared with the traditional classification methods, the deep learning method has obvious advantages in complex scenes(WHU-Hi-HanChuan and WHU-Hi-HongHu), and the classification performance is significantly improved with the increase of training samples. Besides, the accuracy of fully convolutional neural network FPGA is greatly improved comparing to spatial patching-based classification methods. In general, the labeling of hyperspectral data is extremely difficult, and it is still a challenging task to classify the WHU-Hi dataset with a small number of training samples.

## 4. Conclusion

In this article, the WHU-Hi dataset, a new open-source UAV-borne $H^2$ dataset, is built for high spatial resolution hyperspectral image classification. Meanwhile, we have established a benchmark for WHU-Hi dataset with some sart-of-art HSI classification method.Our future research work will pay attention to the practical application of precise crop classification based on UAV-borne $H^2$ imagery.

## Acknowledgements

The authors would like to thank the editor, associate editor, and anonymous reviewers for their helpful comments and advice. This work was supported by National Key Research and Development Program of China under Grant No. 2017YFB0504202, National Natural Science Foundation of China under Grant Nos. 41771385, 41820104006 and 61871299, and by the China Postdoctoral Science Foundation. The authors would also like to give special thanks to the NBL Imaging System Ltd., for its assistance during the WHU-Hi dataset collection, and the WHU-Hi dataset has been published as a benchmark dataset on

http://rsidea.whu.edu.cn/resource_WHUHi_sharing.htm.

## References

[1] Y. Zhong *et al.*, "Mini-UAV-borne hyperspectral remote sensing: from observation and processing to applications," *IEEE Geosci. Remote Sens. Mag.,* vol. 6, no. 4, pp. 46-62, Dec. 2018.

[2] T. Adão *et al.*, "Hyperspectral imaging: A review on UAV-based sensors, data processing and applications for agriculture and forestry," *Remote Sens.,* vol. 9, no. 11, pp. 1110, Nov. 2017.

[3] Y. Zhong, X. Hu, C. Luo, X. Wang, J. Zhao, and L. Zhang, "WHU-Hi: UAV-borne hyperspdectral with high spatial resolution (H2) benchmark datasets and classifier for precise crop identification based on deep convolutional neural network with CRF," *Remote Sens. Environ.,* vol. 250, pp. 112012, 2020.

[4] Y. Zhong, X. Lin, and L. Zhang, "A support vector conditional random fields classifier with a Mahalanobis distance boundary constraint for high spatial resolution remote sensing imagery," *IEEE journal of selected topics in applied earth observations and remote sensing,* vol. 7, no. 4, pp. 1314-1330, 2014.

[5] X. Mei *et al.*, "Spectral-Spatial Attention Networks for Hyperspectral Image Classification," *Remote Sens.,* vol. 11, no. 8, p. 963, Apr. 2019.

[6] Z. Zhong, J. Li, Z. Luo, and M. Chapman, "Spectral–spatial residual network for hyperspectral image classification: A 3-D deep learning framework," *IEEE Trans. Geosci. Remote Sens.,* vol. 56, no. 2, pp. 847-858, Feb. 2018.

[7] M. E. Paoletti, J. M. Haut, R. Fernandez-Beltran, J. Plaza, A. J. Plaza, and F. Pla, "Deep pyramidal residual networks for spectral–spatial hyperspectral image classification," *IEEE Trans. Geosci. Remote Sens.,* vol. 57, no. 2, pp. 740-754, Feb. 2018.

[8] Y. Xu, L. Zhang, B. Du, and F. Zhang, "Spectral-spatial unified networks for hyperspectral image classification," *IEEE Trans. Geosci. Remote Sens. ,* no. 99, pp. 1-17, Feb. 2018.

[9] Z. Zheng, Y. Zhong, A. Ma, and L. Zhang, "FPGA: Fast Patch-Free Global Learning Framework for Fully End-to-End Hyperspectral Image Classification," *IEEE Trans. Geosci. Remote Sens. ,* 2020.